\documentclass[graybox]{svmult}

\usepackage{type1cm}        
\usepackage{makeidx}         
\usepackage{graphicx}        
\usepackage{multicol}        
\usepackage[bottom]{footmisc}

\usepackage{newtxtext}       %
\usepackage{newtxmath}       

\makeindex             

\usepackage[numbers, compress, sort]{natbib}
\usepackage{amsmath}
\usepackage{amssymb}
\usepackage{bbm}

\usepackage{enumerate}

\usepackage{units}

\usepackage{color}
\usepackage{url}

\usepackage[colorlinks]{hyperref}
\hypersetup{%
        plainpages=true,
        breaklinks=true,
        hypertexnames=false,
        pageanchor=true,
        colorlinks=true,
        linkcolor={blue},
        citecolor={magenta},
        urlcolor={blue},
        anchorcolor={black}
      }

\usepackage{mleftright} 

\newcommand{\figref}[1]{\mbox{Fig.~\ref{#1}}}

\newcommand{\secref}[1]{\mbox{Sec.~\ref{#1}}}

\renewcommand{\eqref}[1]{\mbox{Eq.~(\ref{#1})}}

\newcommand{\rd}{\ensuremath{\mathrm{d}}}

\newcommand{\ket}[1]{\mleft|#1\mright\rangle}

\newcommand{\ketbra}[2]{\mleft| #1 \rangle \langle #2 \mright|}

\newcommand{\comm}[2]{\mleft[ #1, #2 \mright]}
\newcommand{\lind}[1]{\mathcal{D}\mleft[#1\mright]}

\newcommand{\sz}{\sigma_z}

\newcommand{\sm}{\sigma_-}
\renewcommand{\sp}{\sigma_+}

\newcommand{\abs}[1]{\mleft|#1\mright|}

\newcommand{\abssq}[1]{\mleft| #1 \mright|^2}

\newcommand{\nn}{\nonumber}

\newcommand{\be}{\begin{equation}}
\newcommand{\ee}{\end{equation}}
\newcommand{\bea}{\begin{eqnarray}}
\newcommand{\eea}{\end{eqnarray}}

\begin{document}

\title*{Quantum optics with giant atoms -- the first five years}

\author{Anton Frisk Kockum}
\institute{Anton Frisk Kockum \at Wallenberg Centre for Quantum Technology, Department of Microtechnology and Nanoscience, Chalmers University of Technology, 412 96 Gothenburg, Sweden, \email{anton.frisk.kockum@chalmers.se}}

\maketitle


\abstract{
In quantum optics, it is common to assume that atoms can be approximated as point-like compared to the wavelength of the light they interact with. However, recent advances in experiments with artificial atoms built from superconducting circuits have shown that this assumption can be violated. Instead, these artificial atoms can couple to an electromagnetic field at multiple points, which are spaced wavelength distances apart. In this chapter, we present a survey of such systems, which we call \textit{giant atoms}. The main novelty of giant atoms is that the multiple coupling points give rise to interference effects that are not present in quantum optics with ordinary, small atoms. We discuss both theoretical and experimental results for single and multiple giant atoms, and show how the interference effects can be used for interesting applications. We also give an outlook for this emerging field of quantum optics.
}



\section{Introduction}
\label{sec:Intro}

Natural atoms are so small (radius $r \approx \unit[10^{-10}]{m}$) that they can be considered point-like when they interact with light at optical frequencies (wavelength $\lambda \approx \unit[10^{-6} - 10^{-7}]{m}$)~\cite{Leibfried2003}. If the atoms are excited to high Rydberg states, they can reach larger sizes ($r \approx \unit[10^{-8} - 10^{-7}]{m}$), but quantum-optics experiments with such atoms have them interact with microwave radiation, which has much longer wavelength ($\lambda \approx \unit[10^{-2} - 10^{-1}]{m}$)~\cite{Haroche2013}. It has thus been well justified in theoretical treatments of quantum optics to assume $r \ll \lambda$, called the \textit{dipole approximation}, which simplifies the description of the interaction between light and matter~\cite{Walls2008}.

In recent years, experimental investigations of quantum optics have expanded to systems with \textit{artificial atoms}, i.e., engineered quantum systems such as quantum dots~\cite{Hanson2007} and superconducting quantum bits (qubits)~\cite{You2011, Xiang2013, Gu2017, Kockum2019a}, which emulate essential aspects of natural atoms. The circuits making up superconducting qubits can be large, reaching sizes up to $r \approx \unit[10^{-4} - 10^{-3}]{m}$, but this is still small when compared with the wavelength of the microwave fields they interact with.

In 2014, one experiment~\cite{Gustafsson2014} forced quantum opticians to reconsider the dipole approximation. In that experiment, a superconducting transmon qubit~\cite{Koch2007} was coupled to surface acoustic waves (SAWs)~\cite{Datta1986, Morgan2007}. Due to the low propagation velocity of SAWs, their wavelength was $\lambda \approx \unit[10^{-6}]{m}$, and the qubit, due to its layout with an interdigitated capacitance, coupled to the SAWs at multiple points, which were spaced $\lambda/4$ apart.

Motivated by this experiment, theoretical investigations on \textit{giant atoms} were initiated~\cite{Kockum2014}. The main finding was that the multiple coupling points lead to interference effects, e.g., the coupling of the giant atom to its environment becomes frequency-dependent~\cite{Kockum2014}.

These initial experimental and theoretical works on giant atoms were published five years ago, at the time of writing for this book chapter. In this chapter, we give a brief survey of the developments in the field of quantum optics with giant atoms that have followed since. We begin in \secref{sec:Theory} with theory for giant atoms, looking first at the properties of a single giant atom (\secref{sec:OneGiantAtom}), including what happens when the coupling points are extremely far apart (\secref{sec:OneGiantAtomTimeDelay}), and then at multiple giant atoms (\secref{sec:MultipleGiantAtoms}). In \secref{sec:Exp}, we survey the different experimental systems where giant atoms have been implemented or proposed. We conclude with an outlook (\secref{sec:Outlook}) for future work on giant atoms, pointing to several areas where interesting results can be expected.


\section{Theory for giant atoms}
\label{sec:Theory}

The experimental setup where giant atoms were first implemented~\cite{Gustafsson2014} falls into the category of waveguide quantum electrodynamics (QED). In waveguide QED~\cite{Gu2017, Roy2017}, a continuum of bosonic modes can propagate in a one-dimensional (1D) waveguide and interact with atoms coupled to this waveguide. As reviewed in Refs.~\cite{Gu2017, Roy2017}, there is an abundance of theoretical papers dealing with one, two, or more atoms coupled to a 1D waveguide, but they almost all assume that the dipole approximation is valid, or, in other words, that the atoms are ``small''.

\begin{figure}
\centering
\includegraphics[width=\linewidth]{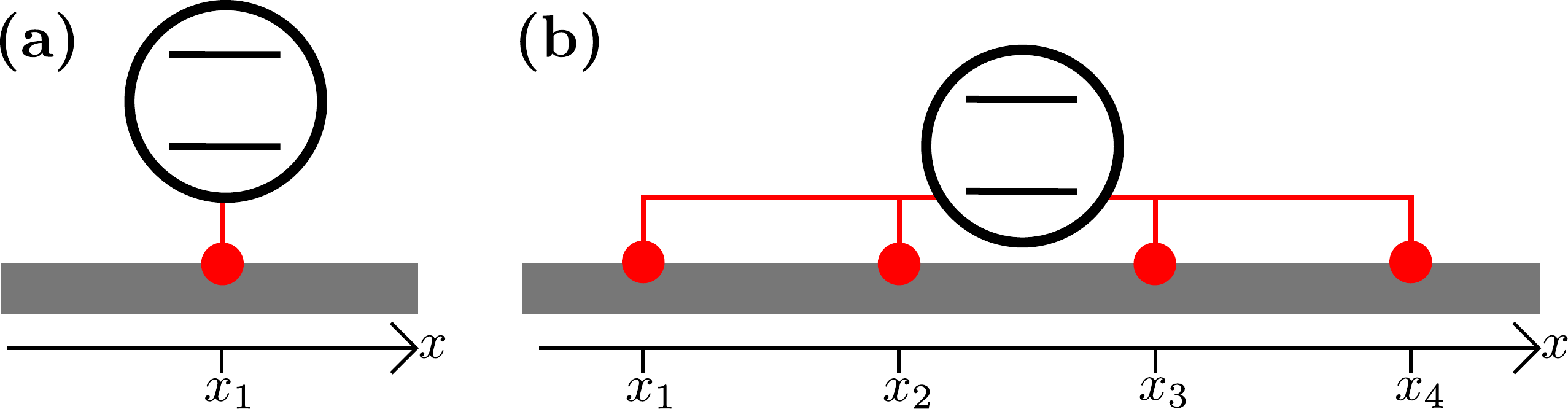}
\caption{The difference between a small atom and a giant atom. 
(a) A small atom (two levels) couples to the 1D waveguide (grey) at a single point (red, coordinate $x_1$).
(b) A giant atom couples to the waveguide at multiple points (labelled $k$, coordinates $x_k$). The distance between two coupling points $k$ and $n$, $\abs{x_k - x_n}$, is \textit{not} negligible compared to the wavelength of the modes in the waveguide that the atom interacts with.
\label{fig:GiantAtomSmallAtom}}
\end{figure}

The difference between small and giant atoms is illustrated in \figref{fig:GiantAtomSmallAtom}. While a small atom, because of its diminutive extent, can be described as being connected to the waveguide at a single point, a giant atom couples to the waveguide at multiple points, and the distance between these points \textit{cannot} be neglected in comparison to the wavelength of the modes in the waveguide that couple to the atom. The relevant wavelength $\lambda$ to compare with is set by the (angular) transition frequency $\omega_{\rm a}$ of the atom and the propagation velocity $v$ in the waveguide: $\lambda = 2\pi v / \omega_{\rm a}$.


\subsection{One giant atom}
\label{sec:OneGiantAtom}

Quantum optics with a single giant atom was first studied theoretically in Ref.~\cite{Kockum2014}, prompted by the experiment in Ref.~\cite{Gustafsson2014} (discussed in \secref{sec:QubitsSAWs}). For a small atom coupled to a continuum of modes, like in \figref{fig:GiantAtomSmallAtom}(a), standard quantum optics procedure is to derive a master equation by assuming that the coupling to the modes is relatively weak and tracing out the modes~\cite{Carmichael1999, Gardiner2004, Walls2008}. When considering whether the same procedure can be applied to a giant atom, there is a new timescale to take into account: the time it takes to travel in the waveguide between coupling points. In Ref.~\cite{Kockum2014}, this time was assumed small compared to the time it takes for an excitation in the atom to relax into the waveguide. With this assumption, the system is \textit{Markovian}, i.e., the time evolution of the atom only depends on the present state of the system, not on the past (for the non-Markovian case, see \secref{sec:OneGiantAtomTimeDelay}). Thus, the standard master-equation derivation from quantum optics with small atoms can be applied here as well.


\subsubsection{Master equation for a giant atom}

The derivation of a master equation for a giant atom starts from the total system Hamiltonian (we use units where $\hbar = 1$ throughout this chapter),
\be
H = H_{\rm a} + H_{\rm wg} + H_{\rm I},
\ee
with the bare atomic Hamiltonian
\be
H_{\rm a} = \sum_m \omega_m \ketbra{m}{m},
\ee
the bare waveguide Hamiltonian
\be
H_{\rm wg} = \sum_j \omega_j \mleft( a^\dag_{{\rm R}j} a_{{\rm R}j} + a^\dag_{{\rm L}j} a_{{\rm L}j}  \mright),
\ee
and the interaction Hamiltonian
\bea
H_{\rm I} &=& \sum_{j, k, m} g_{jkm} \mleft( \sigma_-^{(m)} + \sigma_+^{(m)} \mright) \nn\\
&& \times \mleft( a_{{\rm R}j} e^{- i \omega_j x_k / v} + a_{{\rm L}j} e^{i \omega_j x_k / v} + a^\dag_{{\rm R}j} e^{i \omega_j x_k / v} + a^\dag_{{\rm L}j} e^{- i \omega_j x_k / v} \mright).
\label{eq:Hint}
\eea
Here, the atomic levels are labelled $m = 0, 1, 2, \ldots$, have energies $\omega_m$, and are connected through lowering and raising operators $\sigma_-^{(m)} = \ketbra{m}{m+1}$ and $\sigma_+^{(m)} = \ketbra{m+1}{m}$. The bosonic modes in the waveguide are labelled with indices $j$ and with an index R (L) for right-moving (left-moving) modes. The corresponding annihilation and creation operators are $a$ and $a^\dag$, respectively. The difference to the case of a small atom is the sum over coupling points labelled by $k$ in \eqref{eq:Hint}. The phase factors $e^{\pm i \omega_j x_k / v}$ are not present for a small atom. These phase factors give rise to interference effects. Note that the coupling strengths $g_{jkm}$ can depend on both $j$, $k$, and $m$. 

Following the standard master-equation derivation using the Born-Markov approximation, the resulting master equation becomes
\be
\dot{\rho} = - i \comm{\sum_m \mleft( \omega_m + \Delta_m \mright) \ketbra{m}{m}}{\rho} + \sum_m \Gamma_{m+1, m} \lind{\sigma_-^{(m)}} \rho,
\ee
where $\rho$ is the density matrix for the atom, $\lind{X} \rho = X \rho X^\dag - \frac{1}{2} X^\dag X \rho - \frac{1}{2} \rho X^\dag X$ is the Lindblad super-operator describing relaxation~\cite{Lindblad1976}, and we have assumed negligible temperature $T$, i.e., $\omega_m \gg k_B T$. The relaxation rates for the atomic transitions $\ket{m+1} \rightarrow \ket{m}$ are
\be
\Gamma_{m+1, m} = 4 \pi J \mleft( \omega_{m+1, m} \mright) \abssq{A_m \mleft( \omega_{m+1, m} \mright)},
\ee
where $\omega_{a, b} = \omega_a - \omega_b$, $J (\omega)$ is the density of states at frequency $\omega$ in the waveguide, and we have defined
\be
A_m \mleft( \omega_{j} \mright) = \sum_k g_{jkm} e^{i \omega_j x_k / v}.
\label{eq:A}
\ee
The frequency shifts $\Delta_m$ of the atomic energy levels are Lamb shifts~\cite{Lamb1947, Bethe1947} given by
\be
\Delta_m = 2 \mathcal{P} \int_0^\infty \rd \omega \frac{J(\omega)}{\omega} \mleft( \frac{\abssq{A_m(\omega)} \omega_{m+1,m}}{\omega + \omega_{m+1, m}} - \frac{\abssq{A_{m-1}(\omega)} \omega_{m, m-1}}{\omega - \omega_{m,m-1}} \mright).
\label{eq:Lamb}
\ee

Both the relaxation rates and the Lamb shifts acquire a strong dependence on the atomic transition frequencies, encoded in the factor $A_m \mleft( \omega_{j} \mright)$. For the case of a small atom, $A_m \mleft( \omega_{j} \mright) = g_{jm}$, which is a constant provided that $g_{jm}$ does not depend strongly on $j$. The effect of this frequency dependence for giant atoms can be seen clearly if one considers the simple case of an atom with two coupling points $x_1$ and $x_2$ [compare \figref{fig:GiantAtomSmallAtom}(b)] having equally strong coupling to the waveguide. If the two points are half a wavelength apart, i.e., $\abs{x_1 - x_2} = \pi v / \omega_{m+1,m}$, there will be destructive interference between emission from the two points, and the relaxation for the corresponding atomic transition is completely suppressed: $\Gamma_{m+1, m} = 0$. If the two points are one wavelength apart, there is instead constructive interference and the relaxation rate is enhanced.


\subsubsection{Frequency-dependent relaxation rate}
\label{sec:FrequencyDependentCoupling}

To further understand the frequency-dependence of the relaxation rates and the Lamb shifts, consider the case of a two-level atom coupled to the waveguide at $N$ equidistant points with equal coupling strength at each point. In this case, introducing the notation $\varphi = \omega_{1,0} (x_2 - x_1) / v$, we obtain~\cite{Kockum2014}
\bea
\Gamma_{1,0} &=& \gamma \frac{\sin^2 \mleft( \frac{N}{2} \varphi \mright)}{\sin \mleft( \frac{1}{2} \varphi \mright)} = \gamma \frac{1 - \cos \mleft( N \varphi \mright)}{1 - \cos \mleft( \varphi \mright)},
\label{eq:RelaxationSymmetric} \\
\Delta_1 &=& \gamma \frac{N \sin \mleft( \varphi \mright) - \sin \mleft( N \varphi \mright)}{2 \mleft[ 1 - \cos \mleft( \varphi \mright) \mright]},
\label{eq:LambSymmetric}
\eea
where $\gamma$ is the relaxation rate that the atom would have had if it was coupled to the waveguide only at a single point. To obtain the Lamb shift, we have also made the simplifying assumption that $J (\omega)$ is constant, that the lower limit of the integral in \eqref{eq:Lamb} can be extended down to $-\infty$, and that only the dominating second term in that integral contributes. Since $\Delta_0 = 0$ with these assumptions, \eqref{eq:LambSymmetric} gives the full frequency shift for the two-level atom. In fact, the frequency shift and the Lamb shift are related through a Hilbert transform due to Kramers-Kronig relations~\cite{Cohen-Tannoudji1998}.

\begin{figure}
\centering
\includegraphics[width=\linewidth]{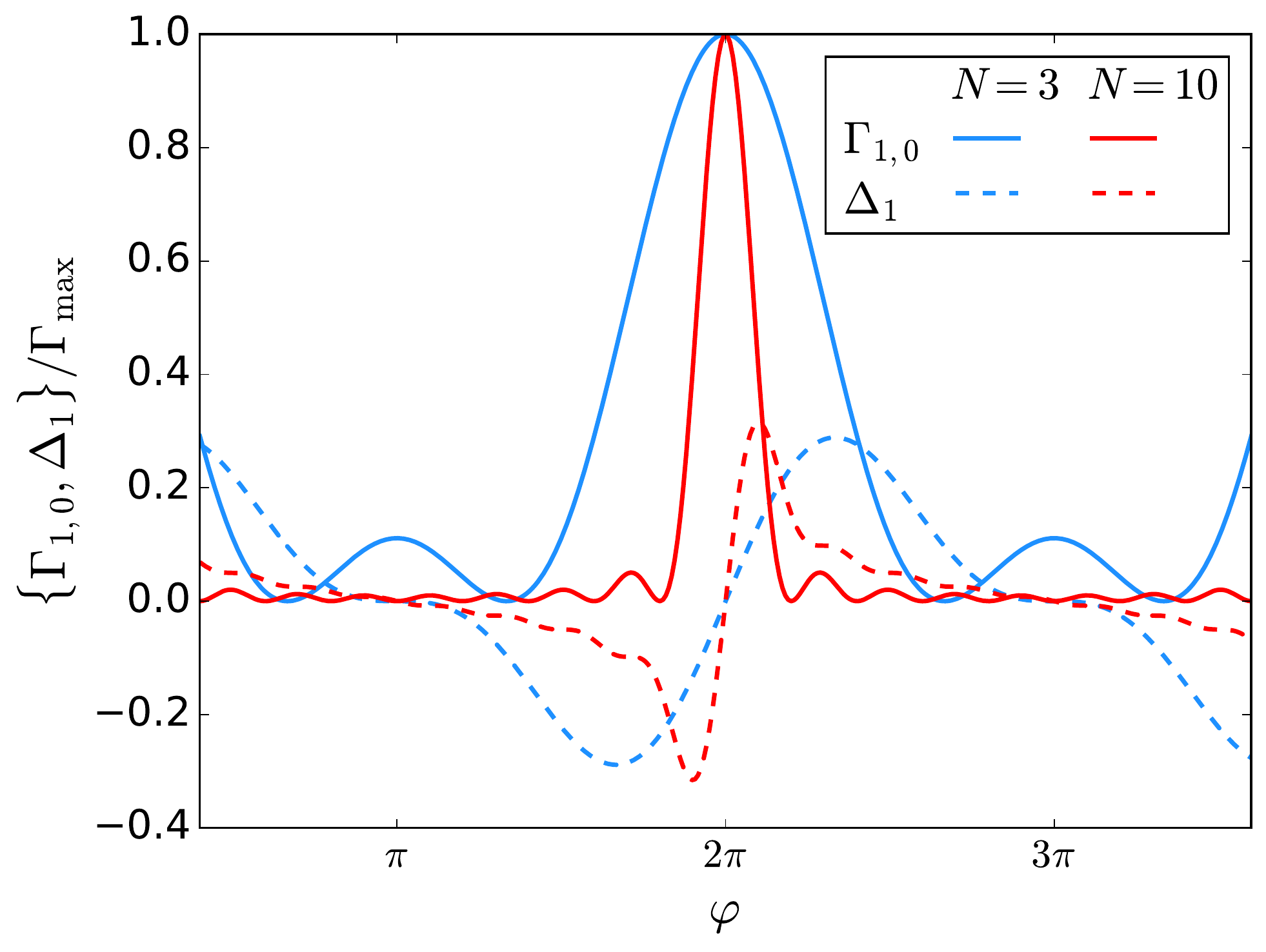}
\caption{Relaxation rates and Lamb shifts for a giant two-level atom with symmetrically spaced coupling points all having the same coupling strength. Red curves: $N=3$ coupling points. Blue curves: $N=10$ coupling points. Solid curves: Relaxation rates $\Gamma_{1,0}$. Dashed curves: Lamb shifts $\Delta_1$. The relaxation rates and Lamb shifts are scaled to the maximum relaxation rate $\Gamma_{\rm max}$ for each $N$. Figure adapted from Ref.~\cite{Kockum2014} with permission.
\label{fig:RelaxationLambSymmetric}}
\end{figure}

The relaxation rates and Lamb shifts in Eqs.~(\ref{eq:RelaxationSymmetric})-(\ref{eq:LambSymmetric}) are plotted for two values of $N$ in \figref{fig:RelaxationLambSymmetric}. The central peak corresponds to the distance between neighbouring coupling points being one wavelength. Note that the frequency dependence becomes sharper when more coupling points are added; in frequency units, the width of the central peak is approximately $\omega_{1,0} / 2 \pi N$. This sharpness can be used to determine when the Markovian approximation underlying the master-equation derivation breaks down, which happens roughly when the relaxation rate changes noticeably within the linewidth of the atom, i.e., when $\Gamma_{1,0} \approx \omega_{1,0} / 2 \pi N$. Interestingly, this is approximately the same condition as when the travelling time between the outermost coupling points, $2 \pi (N - 1) / \omega_{1,0}$, becomes comparable to the relaxation time $1 / \Gamma_{1,0}$.

An attractive feature of giant atoms is that the frequency-dependence of their relaxation rates (and Lamb shifts) can be \textit{designed}~\cite{Kockum2014}. The frequency dependence is directly determined by \eqref{eq:A}, which simply is a discrete Fourier transform of the coupling point coordinates, weighted by the coupling strength in each point. With $N$ coupling points, an experimentalist thus has $2N -1$ knobs to turn (the translational invariance of the setup removes one degree of freedom). With enough coupling points, the curves in \figref{fig:RelaxationLambSymmetric} can be molded into any shape. Note that although the coupling point coordinates and coupling strengths will be fixed in an experiment, superconducting qubits offer the possibility to tune the atomic frequency widely in situ~\cite{Gu2017, Kockum2019a}, making it possible to move between regions with high and low relaxation rates during an experiment.

\begin{figure}
\centering
\includegraphics[width=\linewidth]{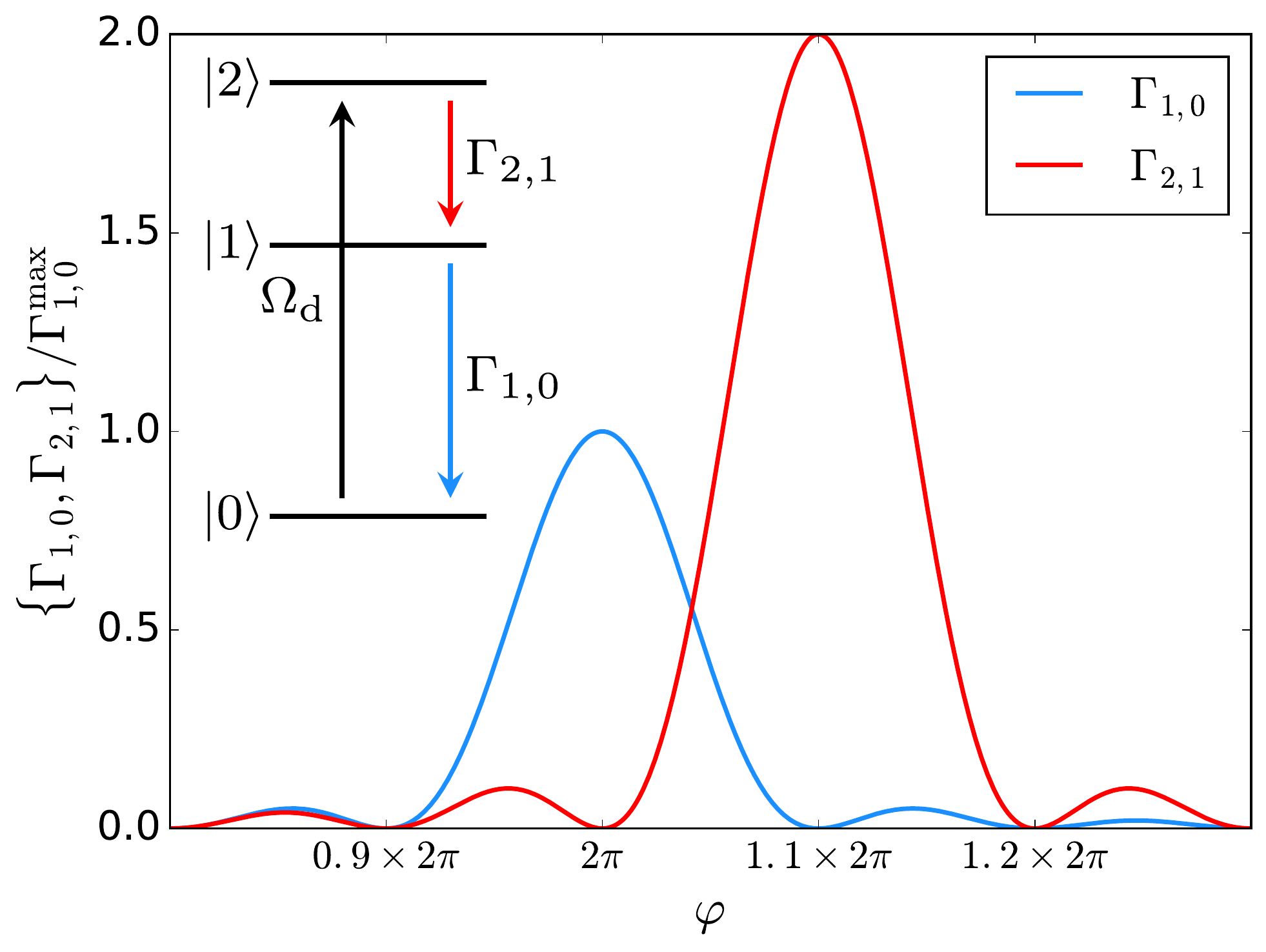}
\caption{Engineering population inversion in a giant atom. The blue curve and the red curve are the relaxation rates $\Gamma_{1,0}$ and $\Gamma_{2,1}$, respectively, as a function of transition frequency $\omega_{1,0}$. The plot assumes $N = 10$ equally spaced coupling points, with equal coupling strengths at all points, and an anharmonicity $\omega_{2,1} - \omega_{1,0} = - 0.1 \times 2 \pi v / (x_2 - x_1)$. The inset shows the level structure with the relaxation rates and a drive of strength $\Omega_{\rm d}$ on the $\ket{0} \leftrightarrow \ket{2}$ transition. Figure adapted from Ref.~\cite{Kockum2014} with permission.
\label{fig:PopulationInversion}}
\end{figure}

If we consider more than two atomic levels, other interesting applications of the frequency-dependent relaxation rate open up. As illustrated in \figref{fig:PopulationInversion}, if the atomic transition frequencies $\omega_{1,0} \neq \omega_{2,1}$, it is possible to engineer the relaxation rates such that $\Gamma_{2,1}$ is at a maximum when $\Gamma_{1,0}$ is at a minimum. At that point, one can then create population inversion, and thus lasing, by driving the transition from $\ket{0}$ to $\ket{2}$~\cite{Kockum2014}. Recent experiments have been making use of this possibility to control the ratio of relaxation rates to enable electromagnetically induced transparency (EIT)~\cite{Andersson2019a, Vadiraj2020}.


\subsubsection{Comparison with an atom in front of a mirror}

It is possible to engineer frequency-dependent relaxation rates and Lamb shifts also for small atoms. This can be achieved by placing a small atom in front of a mirror instead of in an open waveguide, a setup which has been considered in several theoretical~\cite{Meschede1990, Dorner2002, Beige2002, Dong2009, Koshino2012, Wang2012, Tufarelli2013, Fang2015, Shi2015, Pichler2016} and experimental works~\cite{Eschner2001, Wilson2003, Dubin2007, Hoi2015, Wen2018, Wen2019}. Here, the atomic relaxation can be enhanced or suppressed by interference with the mirror image of the atom. This setup is equivalent to a giant atom with two coupling points in a unidirectional waveguide. 

However, this is the limit with a small atom in front of a mirror. In such a setup, it is not possible to increase the number of coupling points, or to have different coupling strengths at different coupling points, which means that the frequency dependence cannot be designed like for a giant atom. Furthermore, since propagation is unidirectional, it is not possible to have more advanced scattering, possible with a giant atom, where both reflection and transmission are influenced by interference between coupling points.


\subsubsection{Coupling a giant atom to a cavity}

By introducing reflective boundary conditions at both ends of the waveguide in \figref{fig:GiantAtomSmallAtom}, a multi-mode cavity will be formed. The coupling of a giant atom to such a cavity has yet to be explored as thoroughly as the open-waveguide case. We can see that similar interference effects as in the open waveguide will come into play. It will thus, for example, be possible to arrange the coupling points such that the giant atom couples strongly to some modes of the cavity and is decoupled from other modes. This can to some extent already be achieved with a small atom, whose single coupling point can be at a node for some modes and at an antinode for others. However, we note that a recent theory proposal~\cite{Ciani2017} uses a superconducting qubit with tunable coupling connected at multiple points to two resonators to cancel certain unwanted interaction terms while keeping desired interaction terms; it is shown that this would not have been possible with a small atom.


\subsection{One giant atom with time delay}
\label{sec:OneGiantAtomTimeDelay}

Consider a giant atom with two coupling points spaced such that it takes a time $\tau$ for light (or sound) to travel between them. In the previous section, it was assumed that $\tau$ was small compared to the relaxation time $1 / \Gamma$. When this no longer is the case, the giant atom enters the non-Markovian regime, where the time evolution of the system can depend on what the system state was at an earlier time. In a giant atom, this non-Markovianity can manifest itself in revivals of the atomic population if energy is sent out from the atom at one coupling point and later is reabsorbed at another coupling point.

Four theoretical studies~\cite{Guo2017, Ask2019, Guo2019, Guo2019a} have explored this regime (the latter three considering more than two coupling points). In Ref.~\cite{Ask2019}, it was shown that a $\Gamma \tau = 1$ constitutes a sharp border for when time-delay effects become visible. When the system transitions from $\Gamma \tau < 1$ to $\Gamma \tau > 1$, the response of the giant atom to a weak coherent probe goes from showing one resonance to showing two. This is similar to the appearance of a vacuum Rabi splitting when an atom becomes strongly coupled to a cavity (the mathematical condition for the appearance of the splitting is actually exactly the same as for an atom in a multi-mode cavity~\cite{Ask2019, Krimer2014}). In the case of the giant atom, the multiple coupling points act as a cavity when the coupling becomes strong enough or the travelling time becomes long enough.

In Ref.~\cite{Guo2017}, the cases $\tau > \Gamma$ and $\tau \gg \Gamma$ were studied in more detail. As $\tau$ increases, an initially excited giant atom exhibits more and more revivals of its population. In the limit of large $\tau$, it turns out that the total energy stored in the giant atom and between its coupling points no longer decays exponentially with time $t$, as for a small atom, but instead decays polynomially ($\propto 1 / \sqrt{t}$). Furthermore, the time scale for this decay is no longer set by the decay rate $\Gamma$, but by the travel time $\tau$. These predictions for a giant atom with time delay were recently confirmed in an experiment~\cite{Andersson2019} (see \secref{sec:QubitsSAWs} for more on the experimental platform used).

In Ref.~\cite{Guo2019}, it was shown that extending the setup from Ref.~\cite{Guo2017} to more three or more coupling points enables qualitatively different phenomena: oscillating bound states. These oscillating bound states do not decay into the waveguide, but the energy oscillates persistently between the atom and the waveguide modes in-between the outermost coupling points of the atom. This result appears connected to that of Ref.~\cite{Ask2019} discussed above, and similar results have been obtained in Ref.~\cite{Guo2019a}.

There are similarities between a giant atom with time delay and the previously studied~\cite{Dorner2002, Tufarelli2013, Pichler2016} setup with a small atom placed far from a mirror. However, in the giant-atom case scattering processes will involve both reflection and transmission, and the second-order correlation functions for these signals, calculated in Ref.~\cite{Guo2017}, exhibit oscillations between bunching and anti-bunching on a timescale set by $\tau$.


\subsection{Multiple giant atoms}
\label{sec:MultipleGiantAtoms}

When multiple small atoms are coupled to a waveguide, they can be spaced wavelength distances apart, which leads to interference effects influencing the collective behaviour of the atoms~\cite{Gu2017, Roy2017, Lehmberg1970a, Lehmberg1970, Lalumiere2013, Zheng2013a}. Well-known examples include super- and sub-radiance~\cite{Dicke1954, Lalumiere2013}, i.e. increased and decreased emission rates due to collective decay, and an effective coupling (sometimes called collective Lamb shift) between pairs of atoms, mediated by virtual photons in the transmission line~\cite{Friedberg1973, Scully2010, Wen2019}. Given this, one might wonder whether there is something left to set multiple giant atoms apart from multiple small atoms. After all, it was mainly the interference effects that separated a single giant atom from a single small atom.

\begin{figure}
\centering
\includegraphics[width=0.75\linewidth]{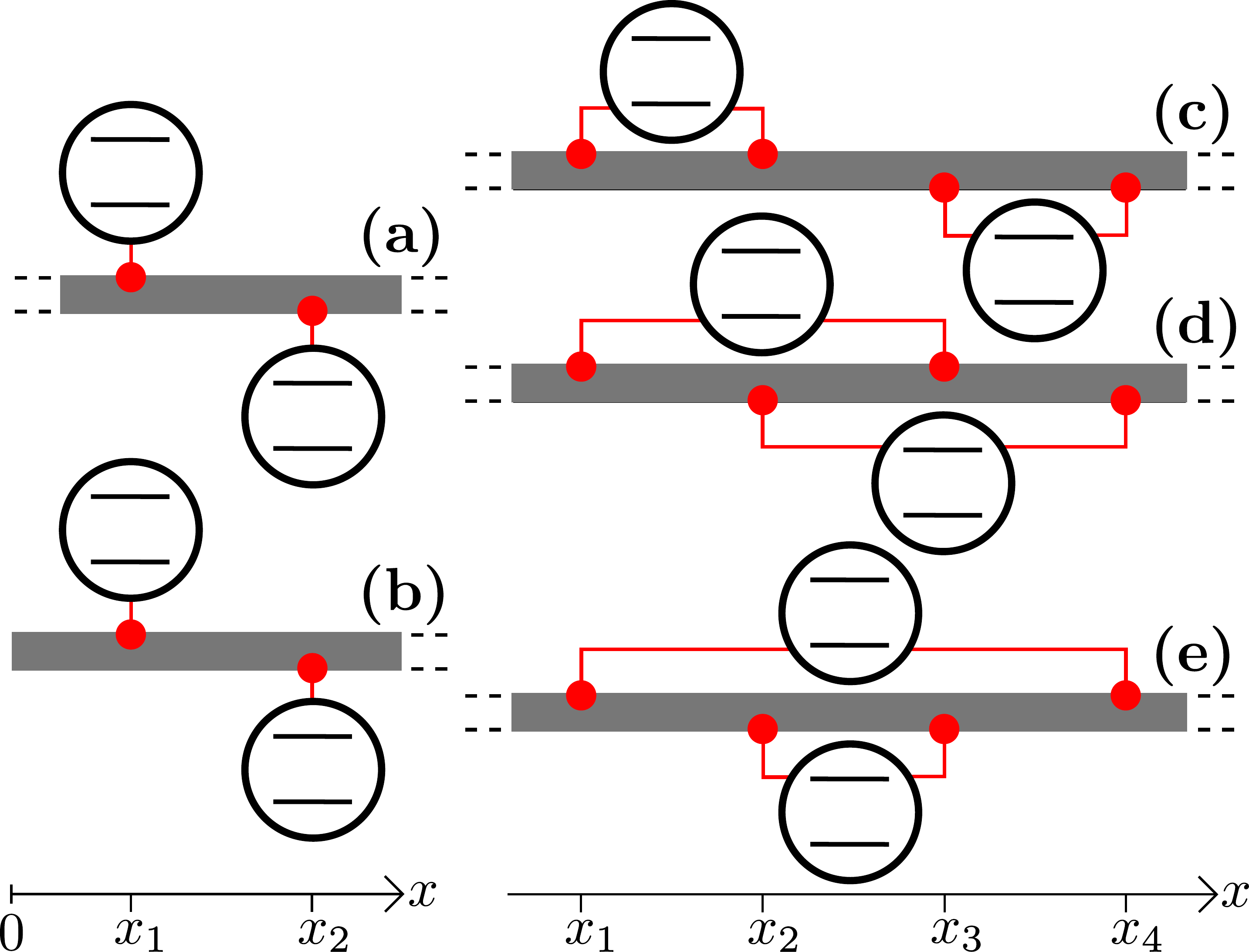}
\caption{Setups for two small and two giant atoms. 
(a) Two small atoms in an open waveguide.
(b) Two small atoms in a waveguide terminated by a mirror on the left.
(c) Two ``separate'' giant atoms, where the rightmost coupling point of the left atom is left of the leftmost coupling point of the right atom.
(d) Two ``braided'' giant atoms, where each atom has a coupling point that lies in-between the two coupling points of the other atom.
(e) Two ``nested'' giant atoms, where the coupling points of one atom all lie in-between the coupling points of the other atom.
Figure adapted from Ref.~\cite{Kockum2018} with permission.
\label{fig:2AtomSetups}}
\end{figure}

In Ref.~\cite{Kockum2018}, the properties of multiple giant atoms was studied thoroughly and compared to those of multiple small atoms. The simplest cases considered are pictured in \figref{fig:2AtomSetups}. For each of these setups, a master equation of the same form can be derived, assuming again that the travel time between coupling points is negligible:
\bea
\dot{\rho} &=& - i \comm{\omega_a' \frac{\sz^a}{2} + \omega_b' \frac{\sz^b}{2} + g \mleft( \sm^a \sp^b + \sp^a \sm^b \mright)}{\rho} \nn\\
&&+ \Gamma_a \lind{\sm^a}\rho +  \Gamma_b \lind{\sm^b}\rho + \Gamma_{\rm coll} \mleft[ \mleft( \sm^a \rho \sp^b - \frac{1}{2}\mleft\{ \sp^a \sm^b , \rho \mright\} \mright) + \text{H.c.} \mright], \quad
\label{eq:ME2Atoms}
\eea
where $\omega_j'$ is the transition frequency of atom $j$ (we label the left atom $a$ and the right atom $b$) including Lamb shifts, $g$ is the strength of the exchange interaction mediated by the waveguide between the atoms, $\Gamma_j$ is the individual relaxation rate of atom $j$, $\Gamma_{\rm coll}$ is the collective relaxation rate, and H.c.~denotes Hermitian conjugate.

\begin{figure}
\centering
\includegraphics[width=\linewidth]{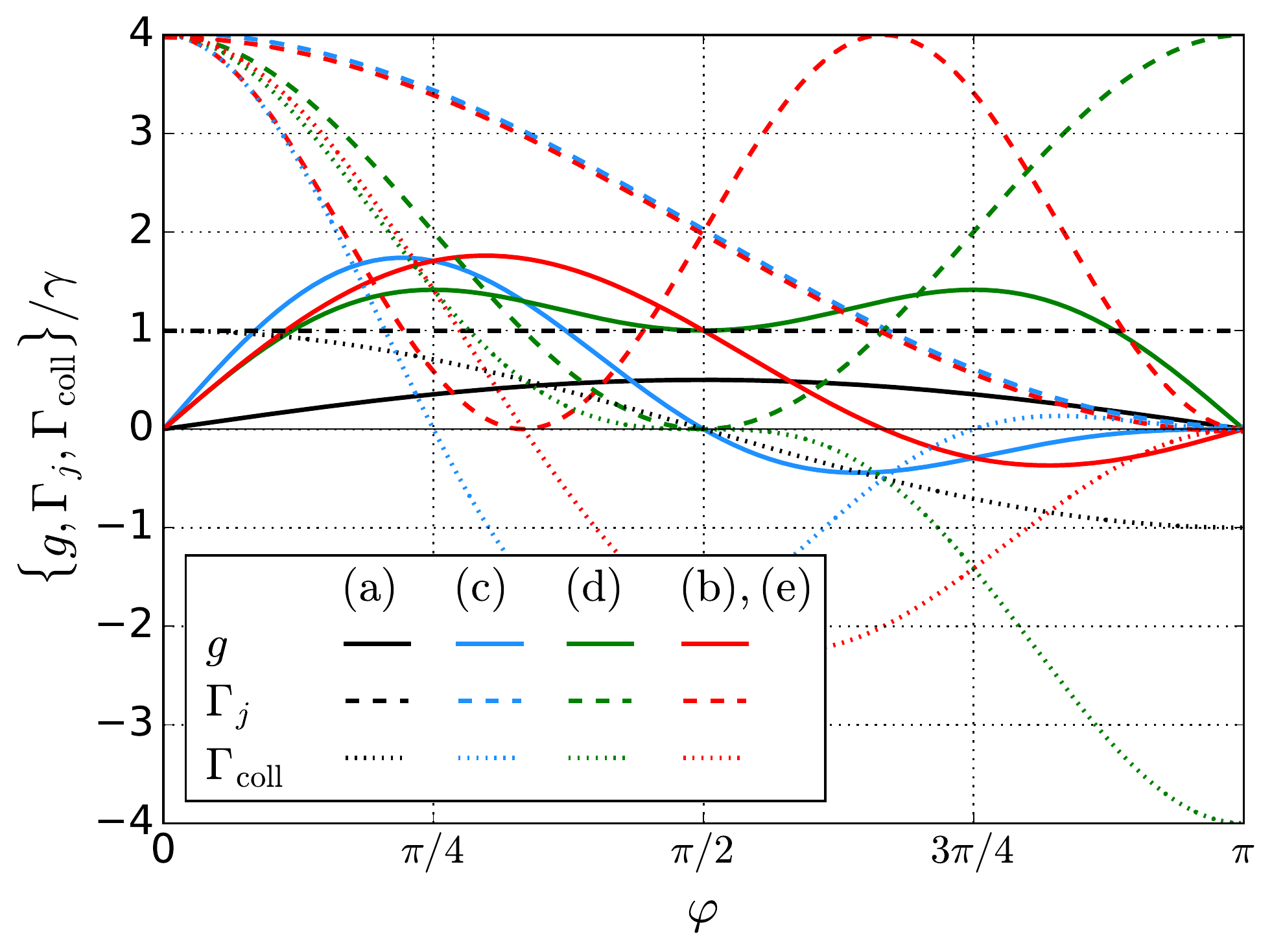}
\caption{Exchange interaction $g$ (solid curves), individual relaxation rates $\Gamma_j$ (dashed curves), and collective relaxation rates $\Gamma_{\rm coll}$ (dotted curves) as a function of $\varphi$ for the setups in \figref{fig:2AtomSetups}. The colours of the curves denote the ordering of coupling points: $ab$ [small atoms, \figref{fig:2AtomSetups}(a), black], $aabb$ [separate giant atoms, \figref{fig:2AtomSetups}(c), blue], $abab$ [braided giant atoms, \figref{fig:2AtomSetups}(d), green], and $abba$ [nested giant atoms, \figref{fig:2AtomSetups}(e), red]. The last case is qualitatively equivalent to small atoms in front of a mirror [\figref{fig:2AtomSetups}(b)]. For this case, there are two dashed curves (red), one for $\Gamma_a$ and one for $\Gamma_b$. Figure adapted from Ref.~\cite{Kockum2018} with permission.
\label{fig:DecoherenceAndCouplingComparisons}}
\end{figure}

Assuming that the atoms couple to the waveguide with equal strength at each coupling point, and that the distances between neighbouring coupling points are equal, corresponding to a phase shift $\varphi$, the coefficients $g$, $\Gamma_j$, and $\Gamma_{\rm coll}$ in \eqref{eq:ME2Atoms} have simple expressions as functions of $\varphi$~\cite{Kockum2018}. These functions are plotted in \figref{fig:DecoherenceAndCouplingComparisons} for all the setups in \figref{fig:2AtomSetups}. Looking at the individual relaxation rates (dashed curves), we see that they are always non-zero for small atoms in an open waveguide, but for setups with giant atoms there are points where $\Gamma_j = 0$, as we know from the discussion of single giant atoms in \secref{sec:OneGiantAtom}. Furthermore, at the points where $\Gamma_j = 0$, the collective relaxation rate $\Gamma_{\rm coll}$ also goes to zero. It is thus clear that setups with multiple giant atoms can be completely protected from relaxation into the waveguide.

The most remarkable feature in \figref{fig:DecoherenceAndCouplingComparisons} is found when looking at the behaviour of the exchange interaction $g$ at the points where the relaxation rates are zero. One might think that since interference effects at these points prevent the atoms from relaxing into the waveguide, it should not be possible for the waveguide to mediate interaction between the atoms. However, it turns out that $g$ can be \textit{non-zero} here for one of the three giant-atom setups: the braided giant atoms. This effect has recently been confirmed in experiment~\cite{Kannan2019} (see \secref{sec:QubitsTL} for more on the experimental platform used).

One way to understand this protected interaction is to note that $\Gamma_j = 0$ when the phase between the coupling points of atom $j$ is an odd integer multiple of $\pi$. The collective relaxation is due to interference between emission from coupling points of different atoms, but the sum total of these contributions is zero if the emissions from the two coupling points of one of the atoms interfere destructively. The exchange interaction arises due to emission from coupling points of one atom being absorbed at coupling points of the other atom. If the giant atoms are in the separate or nested configurations, the emissions from the two coupling points of atom $b$ cancel if $\Gamma_b = 0$, but in the case of \textit{braided} giant atoms, the two inner coupling points are placed in-between the coupling points of the other atom, so there is no condition forcing the contributions from the two coupling points of the other atom to interfere destructively.

We note that the protected interaction with braided giant atoms is reminiscent of the interaction between two small atoms in a waveguide with a bandgap~\cite{Kurizki1990, Lambropoulos2000, Sundaresan2019}. In that case, a bound state of photons forms around each atom that has a frequency in the bandgap, where propagation in the waveguide is impossible. The extension of these bound states decays exponentially with distance, but if two bound states overlap, the atoms can interact without decaying into the waveguide.

It is shown in Ref.~\cite{Kockum2018} that the above conclusions about relations between relaxation rates and exchange interactions in giant atoms remain true even for the most general setups, with an arbitrary number of giant atoms, each having an arbitrary number of coupling points at arbitrary coordinates and with different coupling strength at each coupling point. This opens up interesting possibilities for constructing larger setups with protected exchange interaction between many giant atoms~\cite{Kockum2018}.

It is also interesting that the case of two small atoms in front of a mirror, equivalent to nested giant atoms (red curves in \figref{fig:DecoherenceAndCouplingComparisons}), allows interaction even if one (but not both) of the atoms is prevented from relaxing into the waveguide. This has recently been confirmed in an experiment~\cite{Wen2019} with superconducting qubits in a transmission-line waveguide, and expanded upon in a connected theoretical study~\cite{Lin2019}.

Finally, we note that a recent theoretical study~\cite{Karg2019} extended the treatment from giant atoms to arbitrary quantum systems, e.g., harmonic oscillators, interacting with a waveguide at multiple points. The study took into account losses in the waveguide and also considered the impact of time delays, and showed how these factors can affect the protected interaction that is possible with a nested setup.


\section{Experiments with giant atoms}
\label{sec:Exp}

Waveguide QED can be implemented in several experimental systems~\cite{Gu2017, Roy2017}, e.g., with quantum dots coupled to photonic crystal waveguides~\cite{Arcari2014}, with quantum emitters coupled to plasmons in nanowires~\cite{Akimov2007, Huck2016}, and with natural atoms coupled to optical fibers~\cite{Bajcsy2009}, but the most versatile platform at the moment appears to be superconducting qubits coupled to transmission lines~\cite{Gu2017, Astafiev2010, Astafiev2010a, Hoi2011, Hoi2012, VanLoo2013, Hoi2013a, Hoi2015, Liu2017, Forn-Diaz2017, Wen2018, Mirhosseini2018, Mirhosseini2019, Sundaresan2019, Wen2019, Lu2019}. There are thus many systems where giant atoms could be implemented. So far, as reviewed in this section, experiments have been conducted exclusively with superconducting qubits, coupled to either surface acoustic waves (SAWs, \secref{sec:QubitsSAWs}) or transmission lines (\secref{sec:QubitsTL}). A theoretical proposal exists for an implementation with cold atoms in optical lattices (\secref{sec:ColdAtoms}), and we expect that experiments will eventually be performed using more platforms.


\subsection{Superconducting qubits and surface acoustic waves}
\label{sec:QubitsSAWs}

Superconducting qubits~\cite{You2011, Xiang2013, Gu2017, Kockum2019a} are electrical circuits with capacitances, inductances, and Josephson junctions (which function as nonlinear inductances) that can emulate properties of natural atoms, e.g., energy level structures and coupling to an electromagnetic field. These circuits usually have resonance frequencies $\omega$ on the order of GHz and are cooled to low temperatures $T \ll \hbar \omega / k_{\rm B}$ to prevent that thermal fluctuations interfere with quantum properties.

\begin{figure}
\centering
\includegraphics[width=\linewidth]{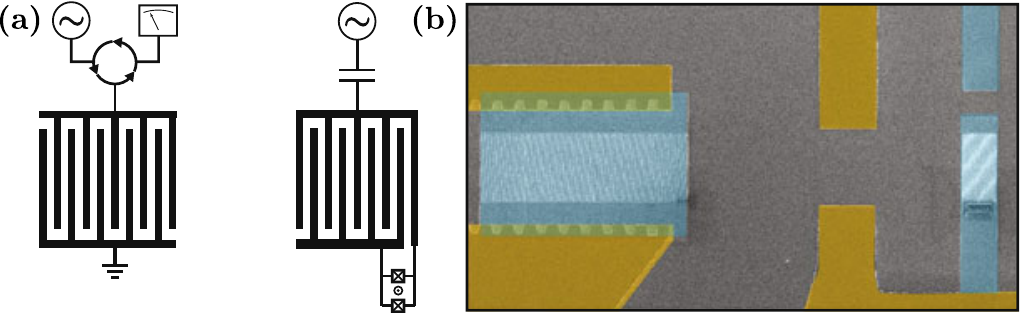}
\caption{Experimental implementation of a giant atom with a superconducting qubit coupled to SAWs.
(a) Sketch of the experimental setup. The IDT on the left is used both to send out SAWs to the right towards the qubit and to convert reflected SAW signals from the qubit into a voltage signal that can be read out. The qubit on the right has its capacitance formed like an IDT to interact with the SAWs. The two islands of the capacitance are also connected through two Josephson junctions (boxes with crosses), which function as a nonlinear inductance, making the qubit essentially an anharmonic $LC$ oscillator. The qubit can also be driven electrically through a gate on the top.
(b) False-colour image of the experimental sample. The blue parts are the IDT to the right and the qubit to the left. The yellow parts are ground planes and the electrodes connecting to the IDT. The aspect ratio of the IDT, with fingers being much longer than they are wide, collimates the SAW beam such that it travels straight toward the qubit (and also in the opposite direction).
Figure from Ref.~\cite{Aref2016} with permission.
\label{fig:ExpSetupTransmonSAW}}
\end{figure}

In 2014, an experiment~\cite{Gustafsson2014} managed to couple a superconducting qubit of the transmon type~\cite{Koch2007} to SAWs, which are vibrations that propagate on the surface of a substrate~\cite{Datta1986, Morgan2007}. The experimental setup is shown in \figref{fig:ExpSetupTransmonSAW}. The substrate on which the SAWs propagate is piezoelectric, which means that the vibrations acquire an electromagnetic component. Vibrations can be induced by applying an oscillating voltage across two electrodes, in the form of an interdigitated transducer (IDT), placed on the surface. If the spacing between fingers in the electrode matches the wavelength of SAWs at the frequency of the applied signal, the induced SAWs add up coherently. Conversely, propagating SAWs that arrive at the transducer induce charge on the fingers such that the vibrations are converted into a voltage signal. The crucial invention in Ref.~\cite{Gustafsson2014} was to let the capacitance in the transmon qubit double as an IDT to mediate a direct coupling between qubit and SAWs. Because of the slow propagation speed of SAWs, $v \approx \unit[3000]{m/s}$, the IDT finger spacing was on the order of $d \approx \unit[1]{\mu m}$ to match the resonance frequency around $\omega \approx \unit[5]{GHz}$. As can be seen in the figure, many fingers were used in the qubit IDT, which corresponded to tens of wavelengths, making this a truly giant atom.

This first experiment with a giant atom could only probe the atom around a single frequency, since the IDT used to convert signals had a narrow bandwidth. The frequency-dependence of the qubit coupling (see \secref{sec:FrequencyDependentCoupling}) could therefore not be tested. However, the experimental platform with SAWs and qubits, called circuit quantum acoustodynamics (QAD)~\cite{Gustafsson2014, Aref2016, Manenti2017} has been adopted in several research groups. In their experiments~\cite{Manenti2017, Noguchi2017, Moores2018, Satzinger2018, Bolgar2018, Sletten2019, Bienfait2019}, the qubit is coupled to a resonator for the SAW modes. Since the resonator is long, it has a narrow free spectral range, and the frequency-dependent coupling of the qubit is evident from how it couples with different strength to different modes. This selective coupling to modes has been used in a clever way to read out the number of phonons in a mode via the qubit~\cite{Sletten2019}.

A particular advantage of the SAWs is that their slow propagation speed makes it possible to engineer a giant atom with a very long distance between coupling points. In the experiment of Ref.~\cite{Andersson2019}, distances exceeding 400 wavelengths were realized, corresponding to $\Gamma \tau \approx 14$, i.e., well in the non-Markovian regime discussed in \secref{sec:OneGiantAtomTimeDelay}.

Another recent experiment~\cite{Andersson2019a} with a superconducting transmon qubit and SAWs used the possibility to engineer the relaxation rates of the first two transitions of the transmon (see \secref{sec:FrequencyDependentCoupling}) to enable EIT. This appears to be the first time that EIT of a propagating mechanical mode has been demonstrated.


\subsection{Superconducting qubits and microwave transmission lines}
\label{sec:QubitsTL}

Superconducting qubits are usually coupled to microwave transmission lines, or $LC$ resonators, instead of SAWs. Also the setup with a transmission line can be used to implement giant atoms, as proposed in Ref.~\cite{Kockum2014}. One simply couples the transmission line to the qubit at one point, meanders the transmission line back and forth on the chip until a wavelength distance has been reached, and then connects the transmission line to the qubit once more. Due to size limitations, this approach will not allow for distances between coupling points on the order of hundreds of wavelengths or more, as is possible with SAWs. However, with the transmission line it is possible to engineer the coupling at each point and the distance between coupling points with great precision, which can be crucial for demonstrating the interference effects that lie at the heart of giant atoms. 

Two recent experiments have followed this approach to implement one~\cite{Vadiraj2020} and two~\cite{Kannan2019} giant atoms. In the experiment with one giant atom, the frequency-dependent coupling shown in \figref{fig:RelaxationLambSymmetric} was measured and the ability to manipulate the relaxation rates in a multi-level atom as in \figref{fig:PopulationInversion} was shown. In the experiment with two giant atoms, the decoherence-free interaction discussed in \secref{sec:MultipleGiantAtoms} was demonstrated. This opens up interesting possibilities for preparing entangled many-body states in waveguide QED with many atoms, which otherwise is difficult due to the dissipation into the waveguide which always is present for small atoms~\cite{Kannan2019}.


\subsection{Cold atoms in optical lattices}
\label{sec:ColdAtoms}

All experiments with giant atoms so far have taken place in 1D geometries at microwave frequencies and used superconducting qubits. A recent theory proposal~\cite{Gonzalez-Tudela2019} shows how giant atoms instead could be implemented in higher dimensions on another platform for quantum-optics simulation: cold atoms in optical lattices. Here, one would use atoms with two internal states, each of which couples to a different optical lattice, realized by counter-propagating lasers. In one state, the atom mimics a photon moving in a lattice; in the other state, the atom mimics an atom trapped in a specific site. By rapidly modulating the relative positions of the two lattices, it is possible to engineer an effective interaction where the atomic state couples to the photonic state at multiple points~\cite{Gonzalez-Tudela2019}. It may be possible to achieve a similar effect with superconducting qubits coupled to several sites in a 2D lattice of superconducting resonators. While such lattices have been analyzed and realized previously~\cite{Koch2010, Houck2012, Underwood2016}, to the best of our knowledge it has not been suggested previously to couple one qubit to several lattice sites in such a setup. 

The proposed setup with cold atoms displays rich physics with the giant atoms coupled to 2D photonic environments that have a band structure. It is possible to construct interference such that a single giant atom relaxes by only emitting its energy in certain directions. It is also possible to decouple giant atoms completely from the environment, but still have them interact by exchange interactions, like in \secref{sec:MultipleGiantAtoms}. While this interference was possible with just two coupling points per atom in 1D, the 2D case requires at least four coupling points.


\section{Conclusion and outlook}
\label{sec:Outlook}

Giant atoms are emerging as a new, interesting field of quantum optics. Following the first experimental realization and theoretical study in 2014, the field has grown quickly in the past five years. Theoretical investigations have been extended from one to multiple giant atoms, from 1D to higher-dimensional environments coupling to the atoms, and from the Markovian to the non-Markovian regime, where time delays between coupling points matter. These investigations have revealed remarkable properties of giant atoms, including frequency-dependent couplings and decoherence-free interactions, which are hard or impossible to realize with small atoms. 

In parallel, the experimental platform for giant atoms, with SAWs coupled to superconducting qubits, has been further developed. There are now also experiments with superconducting qubits coupled to microwave transmission lines, and an experimental platform with cold atoms in optical lattices has been proposed. The experiments have confirmed many of the theoretical predictions, and also contributed with new ideas for applications of giant atoms.

Looking towards the future, we can formulate a long research agenda for giant atoms. At the heart of this agenda is the fact that giant atoms mainly differ from small atoms by the interference effects introduced by the multiple coupling points, which already has been shown to lead to new effects. It therefore seems prudent to revisit many well-known quantum-optics phenomena to see if giant atoms can enhance them or enable new physics. Below, we give a list of such projects:
\begin{itemize}
\item \textbf{Superradiance:} For multiple small atoms coupled to light, it is well known that quantum interference effects can give rise to enhanced light emission, superradiance, where $N$ atoms emit light at an increased rate, proportional to $N^2$~\cite{Dicke1954, Shammah2018}. The reverse process, ``superabsorption'', is also possible~\cite{Higgins2014, Yang2019}, and may be of importance in photosynthesis and future solar cells. It is thus highly relevant to see if giant atoms can enhance superradiance and superabsorption.
\item \textbf{Ultrastrong coupling:} When the strength of the coupling between light and matter starts to approach the bare resonance frequencies in the system, it is called ultrastrong~\cite{Kockum2019, Forn-Diaz2019}. In this regime, the rotating-wave approximation breaks down and the number of excitations in the system is no longer conserved in the absence of drives. For a giant atom ultrastrongly coupled to an open waveguide, it would be interesting to map out the ground state of the system, since results for a small atom indicates that it should contain virtual photons clustered around each connection point~\cite{Sanchez-Burillo2014}. However, ultrastrong coupling with giant atoms comes with several theoretical challenges, which make analytical results hard to achieve. For example, a giant atom with ultrastrong coupling will inevitably be in a regime where the travel time between coupling points is non-negligible~\cite{Ask2019}.
\item \textbf{Generating nonclassical light:} It has recently been shown that coherently driving a small atom in front of a mirror can lead to the generation of nonclassical states of light with a negative Wigner function~\cite{Quijandria2018}. Could a giant atom do the same?
\item \textbf{Matryoshka atoms:} The topology in \figref{fig:2AtomSetups}(e), nested atoms, is reminiscent of a Russian matryoshka doll. Although it does not enable decoherence-free interaction like braided atoms do, it seems to have other interesting properties. If the distance between the coupling points of the outer atom is large, the outer atom could effectively act as a cavity~\cite{Ask2020a}, similar to what two small atoms placed far away on either side of a central atom can do~\cite{Guimond2016a}. Also, preliminary results indicate that two nested giant atoms can emulate EIT in a $\Lambda$ system without any external drive~\cite{Ask2020b}. With many nested giant atoms, the situation is similar to having many atoms in front of a mirror. Thus, for certain inter-coupling-point distances, these giant atoms should be able to combine into fewer effective larger atoms, as can happen in the mirror case~\cite{Lin2019}.
\item \textbf{Chiral quantum optics:} In some waveguide-QED setups with small atoms, it is possible to realize chiral couplings, i.e., that the atoms only couple to one propagation direction in the waveguide~\cite{Lodahl2017}. Although it is not yet clear if this can be implemented in experiments with giant atoms, it seems interesting to study chiral quantum optics with giant atoms theoretically. A related question is whether interference between light propagating in a waveguide, and light taking the ``shortcut'' between two coupling points through a giant atom, can be used to realize an effective chiral coupling. This was recently answered affirmatively for a setup with two atoms that are both directly coupled to each other and each coupled at its own single point to a waveguide ($\sim \lambda / 4$ apart)~\cite{Guimond2019}.
\end{itemize}
%


\begin{acknowledgement}
AFK acknowledges support from the Knut and Alice Wallenberg foundation and from the Swedish Research Council (grant number 2019-03696).
\end{acknowledgement}


\bibliography{ControlForBibstyle,GiantAtomsReferences}

\end{document}